# A Method for Multi-Hop Question Answering on Persian Knowledge Graph

**Arash Ghafouri**

Computer Engineering Department, Iran University of Science and Technology, Tehran, Iran
E-mail: aghafuri@comp.iust.ac.ir

**Mahdi Firouzmandi**

Computer Engineering Department, Iran University of Science and Technology, Tehran, Iran
E-mail: firouzmandi@gmail.com

**Hasan Naderi***

Computer Engineering Department, Iran University of Science and Technology, Tehran, Iran
E-mail: naderi@iust.ac.ir
*Corresponding author

Abstract:

Question answering systems are the latest evolution in information retrieval technology, designed to accept complex queries in natural language and provide accurate answers using both unstructured and structured knowledge sources. Knowledge Graph Question Answering (KGQA) systems fulfill users' information needs by utilizing structured data, representing a vast number of facts as a graph. However, despite significant advancements, major challenges persist in answering multi-hop complex questions, particularly in Persian. One of the main challenges is the accurate understanding and transformation of these multi-hop complex questions into semantically equivalent SPARQL queries, which allows for precise answer retrieval from knowledge graphs. In this study, to address this issue, a dataset of 5,600 Persian multi-hop complex questions was developed, along with their decomposed forms based on the semantic representation of the questions. Following this, Persian language models were trained using this dataset, and an architecture was proposed for answering complex questions using a Persian knowledge graph. Finally, the proposed method was evaluated against similar systems on the PeCoQ dataset. The results demonstrated the superiority of our approach, with an improvement of 12.57% in F1-score and 12.06% in accuracy compared to the best comparable method.

## 1. Introduction

A knowledge graph is a collection of interconnected entities enriched with semantic labels [1], where nodes represent entities and edges depict the relationships between them [2]. Entities can be various types of named and unnamed objects, such as persons, locations, organizations, events, times, concepts, and more [3]. Knowledge graphs have a wide range of applications, including in search engines, natural language processing, question-answering systems, information extraction, and social networks like LinkedIn and Facebook. The first and most comprehensive Persian-language knowledge graph, named "FarsBase," has been introduced, focusing on general knowledge. Given the absence of an effective knowledge graph in the Persian language, "FarsBase" can serve as one of the most crucial resources for natural language question-answering systems [4].

Answering questions based on a knowledge graph (KGQA) is a fundamental task in the field of natural language processing, focusing on providing answers to questions posed in natural language by utilizing the information stored in a knowledge graph. KGQA has gained significant attention due to its pivotal role in various intelligent applications. For instance, systems such as Amazon Alexa, Apple Siri, and Microsoft Cortana are examples of platforms that leverage KGQA to respond to user inquiries [5].

Early research on KGQA primarily focused on answering simple questions containing a single fact. However, addressing complex questions (involving multiple facts) in English still presents challenges, such as difficulties in understanding complex queries, constructing high-quality datasets, and other related issues. In contrast, research on knowledge graph question answering in Persian has been very limited. This study aims to propose a method for answering multi-hop Persian questions using a knowledge graph.

Given the limited efforts in the Persian language domain concerning knowledge graph question answering systems, developing a method for answering multi-hop questions in Persian using a knowledge graph is essential. This need stems from the importance of enhancing intelligent interactions and improving question answering systems in Persian, thereby enabling users to access the required information more accurately and efficiently.

The primary objective of this research is to answer Persian multi-hop questions over a knowledge graph. This entails enabling the system to accurately and efficiently respond to user queries posed in natural language, which require multiple reasoning steps using the information stored in the knowledge graph. Effectively understanding and reasoning through the steps of complex questions enhances the precision and accuracy of the answers. Moreover, by improving the speed of retrieving answers, users can pose complex queries as a single, comprehensive question rather than multiple related simpler ones, and receive specific and precise responses.

Multi-hop questions refer to those that require traversing multiple steps or stages to arrive at the correct answer [5]. As illustrated in Figure 1, to answer the question "Who is the first wife of the television director who executive produced the show 'Khandevaneh'?", two relationships—'executive' and 'wife'—must be navigated sequentially to reach the correct answer. This type of query, which involves a combination of different relationships, is known as a multi-hop complex question.

The rest of the paper is structured as follows: In the next section, we present a comprehensive review of four main approaches to answering knowledge graph questions. Following this, we explain the proposed method, considering a deep understanding of the topic and the challenges raised. We also evaluate the proposed method, demonstrating its effectiveness based on standard criteria in comparison to existing similar works. Finally, the results of this research are discussed, and suggestions for future research are provided.

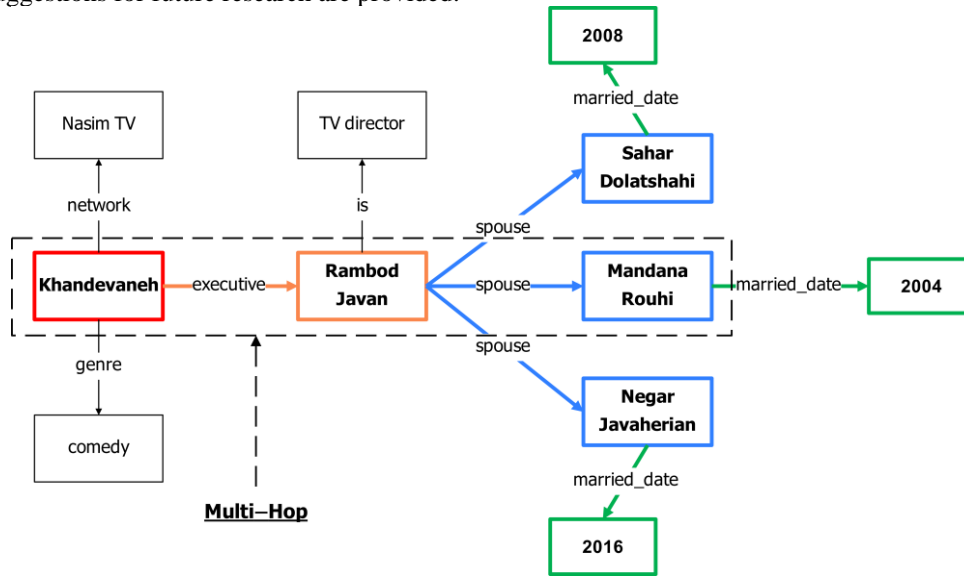

Figure 1: illustrates an example of a multi-hop complex question in KGQA for the query, "Who is the first wife of the television director who executive produced the show 'Khandevaneh'?" We present the related subgraph of the knowledge graph (KG) for this question. The correct path leading to the answer is highlighted with colored borders. The topic entity and the answer entity are shown in bold font and shaded boxes, respectively. The "multi-hop" reasoning process is emphasized within a black dotted box. Different colors are used to indicate various reasoning steps required to reach each entity from the topic entity.

## 2. Related Works

In general, there are four primary approaches to Knowledge Graph Question Answering (KGQA):

**Traditional Approach**: Traditional KBQA methods rely on manually defined templates and rules to parse complex questions. These methods often require linguistic expertise and face scalability issues[6]. Berant et al. [7] implemented a standard bottom-up parser that generates a large set of question templates based on the entities and relationships of a knowledge base and uses a recursive derivation parser to map parsed phrases to the knowledge base entities and relationships using four predefined manual operations. Bast et al. [8] proposed a template-based model called Aqqu, which maps questions based on three templates by identifying entities, matching them to the knowledge base, and selecting the best response template using a ranking model.

**Semantic Parsing (SP) Approach**: In neural semantic parsing-based methods, the goal is to transform natural language questions into executable queries. The natural language question is first understood and interpreted by the system, which then converts it into a general logical form. This logical form is subsequently matched with various knowledge bases, and eventually, it is executed on the relevant knowledge bases according to their specific query formats. In most cases, this executable logical form corresponds to SPARQL queries [9]. Das et al. [10] introduced a neuro-symbolic case-based reasoning (CBR) approach, termed CBR-KBQA, for answering questions in large knowledge bases. CBR-KBQA features a non-parametric memory that stores historical cases, including questions and their corresponding logical forms, alongside a parametric model that generates new logical forms for new questions by retrieving and utilizing relevant cases from the memory. Ye et al. [11] They introduced RnG-KBQA, a Rank-and-Generate approach for KBQA that effectively addresses the coverage issue by incorporating a generation model, while preserving robust generalization capabilities. Shu et al. [12] introduced a KBQA model called TIARA, which utilizes multi-grained retrieval to assist the PLM in focusing on the most relevant knowledge base content, including entities, exemplary logical forms, and schema items. Additionally, constrained decoding is employed to control the output space and minimize generation errors. Yu et al. [13] proposed a new framework called DecAF, which simultaneously generates logical forms and direct answers, then combines the strengths of both to produce the final answer. Cao et al. [14] proposed a program transfer approach that leverages valuable program annotations from resource-rich knowledge bases as external supervisory signals to assist program induction in low-resource knowledge bases that lack such annotations.

**Information Retrieval (IR) Approach**: In information retrieval-based methods, question answering is approached as either a binary classification problem or as the classification of candidate answers. Initially, a distributed representation of the candidate answers and the questions is generated. Finally, the system calculates a score for each candidate answer, and the final answer is determined based on these scores [15]. Ding et al. [16] introduce Evidence Pattern Retrieval (EPR), a method that models the structural dependencies among evidence facts during subgraph extraction. In this approach, dense retrieval is first used to obtain atomic patterns formed by resource pairs when a question is posed. These patterns are then combined to construct candidate evidence patterns, which are scored by a neural model. The highest-scoring pattern is selected to extract the subgraph, facilitating downstream answer reasoning. Saxena et al. [17] propose EmbedKGQA, a method designed to embed knowledge graphs (KG embedding) to address the issue of KG sparsity in multi-hop KGQA tasks. EmbedKGQA proves particularly effective in handling sparse KGs by eliminating the need for answer selection from a pre-defined neighborhood, thereby overcoming a key limitation of previous multi-hop KGQA methods. He et al. [18] introduced a teacher-student framework to address the challenges in multi-hop KBQA. In this framework, the student network is tasked with finding the correct answer, while the teacher network provides intermediate supervision signals by leveraging both forward and backward reasoning. This dual reasoning approach enhances the learning of intermediate entity distributions, resulting in more reliable supervision signals. The method effectively reduces issues related to spurious reasoning and has demonstrated improved reasoning capabilities in the student network across three benchmark datasets. Zhang et al. [19] propose a trainable subgraph retriever (SR) that is independent of the reasoning process, allowing for the enhancement of any subgraph-oriented KBQA model. Jiang et al. [20] proposed UniKGQA, a model that integrates a semantic matching module based on a pre-trained language model (PLM) for question-relation matching, and a matching information propagation module to disseminate this matching information along the directed edges in knowledge graphs.

**Large Language Model Approach**: These methods leverage large language models for question parsing and answering, enhancing the performance of KGQA models. These approaches can be broadly categorized into two main groups: information retrieval-based methods and semantic parsing-based methods [21]. Sen et al. [22] proposed a novel approach for answering complex questions by combining a knowledge graph retriever, based on an end-to-end KGQA model, with a language model that reasons over the retrieved facts. Chakraborty [21] evaluated the ability of large language models (LLMs) to answer questions requiring multi-hop reasoning over knowledge graphs (KGs) and demonstrated that different approaches are necessary for extracting and presenting relevant information to the LLM depending on the size and nature of the KG. The approach was tested on six KGs, both with and without example-specific subgraphs, and it was shown that both IR and SP-based methods can be effectively adopted by LLMs, yielding highly competitive performance.

Figure 2 summarizes the main approaches to question answering from knowledge graphs. These approaches can be categorized into four main types: 1) Traditional methods, which rely on predefined rules and templates for parsing complex questions and generating a logical form. 2) Neural Semantic Parsing, where complex questions are understood, converted into a general logical form, and executed on the knowledge base. 3) Information Retrieval, where a distributed representation of candidate answers and questions is first generated, and the final answer is selected based on the score of each candidate answer. 4) Large Language Models, which leverage powerful language models for question parsing and answering, enhancing the performance of knowledge graph question-answering (KGQA) models.

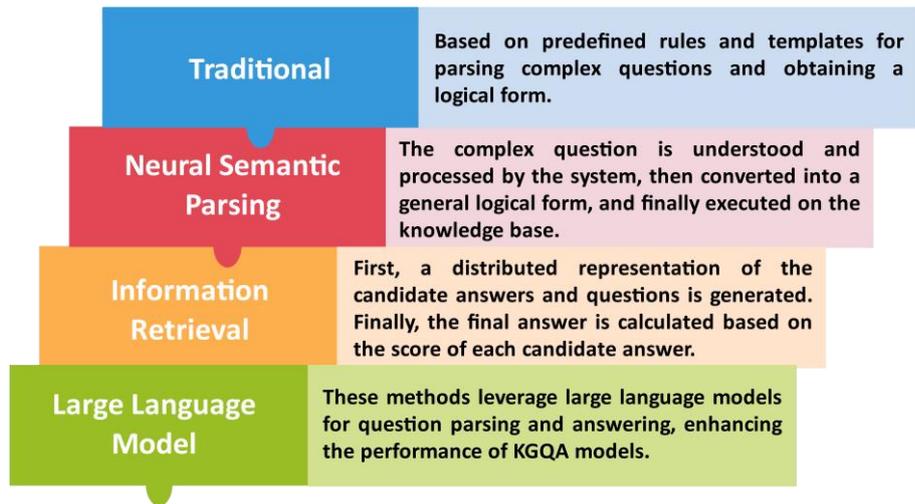

Figure 2. The implementation method involves four main approaches in question answering based on the knowledge graph.

In Persian, only one significant work has been done on question answering using the Persian knowledge graph, FarsBase. Etezadi et al. [23] propose a method that maps the question to its corresponding logical form (SPARQL). This approach focuses on addressing the complexities in the Persian language and generates a set of possible logical forms for the given complex question. Finally, to select the best logical form, the method utilizes Multilingual-BERT.

## 3. Background

A knowledge base is a structured database that contains a collection of facts about entities, typically derived from structured repositories like WIKIDATA [24], FreeBase [25], YAGO [26], DBpedia [27] and others, or extracted from encyclopedias like WIKIPEDIA [28]. Knowledge bases are usually represented as graphs, hence the term "knowledge graphs" is used for knowledge bases stored in a graph structure [29]. Since the advent of the Semantic Web, knowledge graphs have become associated with Linked Data projects, both focusing on linking entities and concepts [30]. Linked Data refers to a type of structured data that is interlinked with other data, making it suitable for semantic queries. Semantic Web and Linked Data are commonly stored in the RDF (Resource Description Framework) data format. RDF is a graph-based model used to describe interconnected entities such as objects, places, events, abstract concepts and the like. This model provides the best framework for integrating, linking, and reusing data to represent and present a knowledge graph. In Figure 3, the RDF data structure is schematically represented. This structure, as a graph-based model, is used to represent entities and their relationships. RDF consists of three core elements: Subject, Predicate, and Object, collectively referred to as a "Triple".

The first and most comprehensive knowledge base dedicated to the Persian language, specifically in the domain of general knowledge, is known as FarsBase [31]. Given the absence of a useful knowledge base in the Persian language, FarsBase can serve as one of the most important resources for natural language question answering [32]. The data in a knowledge graph is typically described in triples, consisting of a subject, predicate and object meaning that the subject and object are connected by a predicate [33].

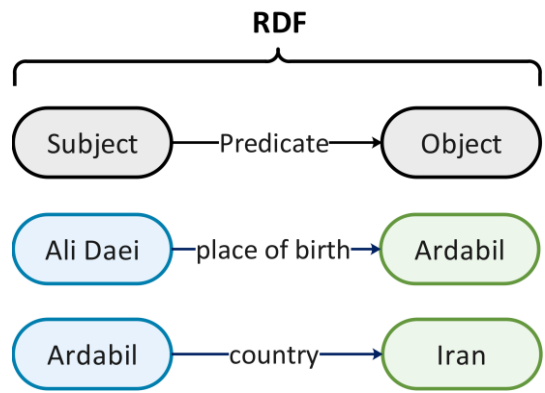

Figure 3. This image illustrates the abstract structure of a knowledge graph, which consists of three components: the subject, predicate, and object, forming a Resource Description Framework (RDF) triple. In the diagram, Ali Daei is shown as the subject, connected to Ardabil as the object through the predicate "place of birth". Additionally, Ardabil, which served as the object in the previous triple, now becomes the subject in the next triple, linked to Iran as the object through the predicate "country".

In the Persian language, there exists only one question-answering dataset based on a knowledge graph. This dataset, known as PeCoQ[34], comprises 10,000 complex questions and their corresponding answers, extracted from the FarsBase knowledge graph. For each question, a SPARQL query and two manually crafted linguistic paraphrases are also provided. The dataset features various types of complexities, including multi-relational, multi-entity, sequential, and temporal constraints.

## 3. Proposed Methods

We propose an innovative method for answering multi-hop complex questions, which consists of four key steps: 1) Decomposition of the complex question into simpler questions, 2) Named Entity Recognition and Linking from the simpler questions, 3) Logical Form Generation for each simple question, and 4) Sequential and step-by-step execution of the generated logical forms on the knowledge graph. The diagram related to this process is presented in Figure 4.

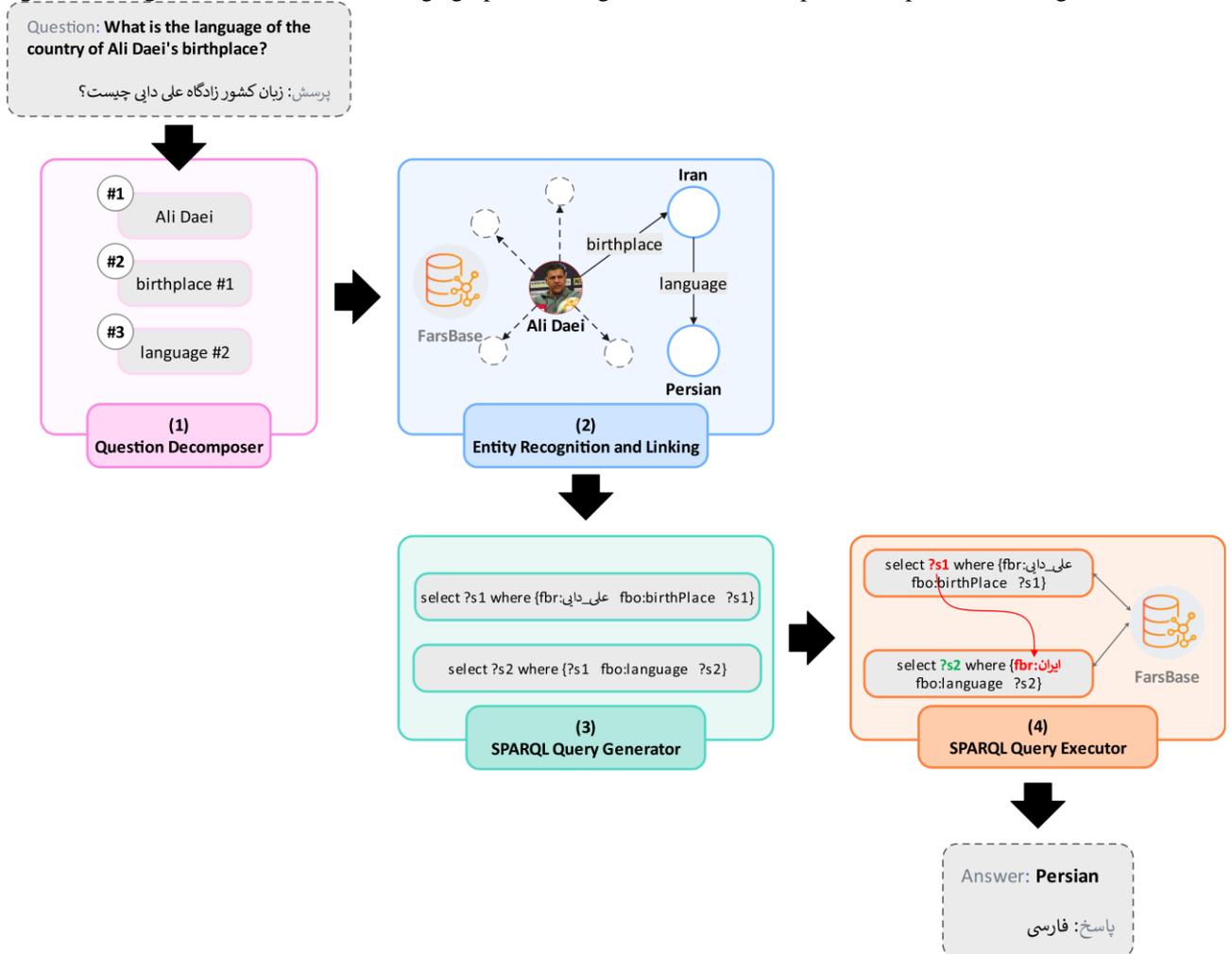

Figure 4. The proposed architecture for answering multi-hop complex questions on the FarsBase knowledge graph. The architecture is composed of four main components: (1) The question decomposer breaks down a complex question into simpler sub-questions, (2) The entity recognition and linking component identifies named entities in the sub-questions and links them to the FarsBase knowledge graph, (3) The SPARQL query generator converts the sub-questions into corresponding SPARQL queries, and (4) The SPARQL query executor sequentially executes the generated queries on FarsBase, retrieving the final answer.

Considering the architecture presented in Fig. 4, the complex question is initially processed by the question decomposition component. This component breaks down the complex question into smaller, semantically meaningful segments, which we refer to as the Meaning Representations of Decomposed Complex Persian Questions (MRDCPQs). These decomposed segments are then sent to the named entity recognition component, which identifies the named entities in each MRDCPQ. After recognition, the named entities are passed to the entity linking component, which links the identified entities to their corresponding entries in the FarsBase knowledge graph. Next, each MRDCPQ, enriched with the extracted information, is converted into a SPARQL query by the SPARQL query generation component. These generated SPARQL queries are then sequentially executed by the query execution component on the FarsBase knowledge graph, where the final answer is extracted. In the following, we will describe in detail the function of each component in the proposed method's architecture.

## 3.1 Question Decomposition Component

Understanding complex questions necessitates multi-hop reasoning. To enhance the comprehensibility of this reasoning and make it easier for users to engage with and comprehend it, this component decomposes complex questions into smaller, semantically meaningful segments. Our approach draws inspiration from the method proposed by Wolfson et al. [35], which emphasizes breaking down complex questions into simpler sub-components. The question decomposition process performed by this component is demonstrated in Figure 5 below.

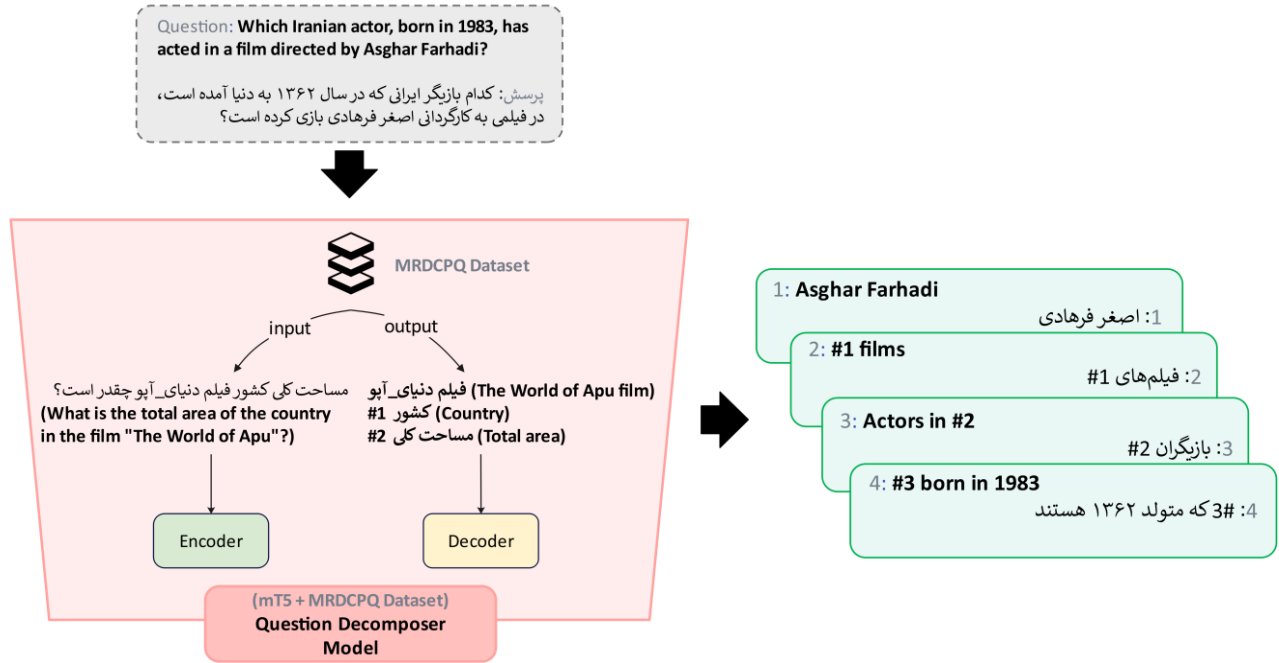

Figure 5. The Question Decomposition Component breaks down complex Persian questions into smaller, semantically meaningful sub-questions. The illustrated model leverages the MRDCPQ dataset, utilizing a pre-trained mT5 encoder-decoder architecture to decompose input questions for easier multi-hop reasoning. This process simplifies complex questions into manageable semantic representations, as shown in the example involving a question about an Iranian actor from Asghar Farhadi's films..

### 3.1.1 MRDCPQ Dataset

In the absence of a ready-made dataset for decomposing complex questions into simpler ones in Persian, we have created the Persian MRDCPQ dataset as the first dataset for semantically decomposing complex questions into smaller semantic units. This dataset has been created according to the standards of the BREAK [36] dataset and based on the PeCoQ [34] dataset, which is related to complex Persian questions over the FarsBase [25] knowledge graph. To provide the necessary infrastructure for better analysis and reasoning over complex multi-hop Persian questions, the MRDCPQ dataset was created. This dataset includes 5,600 complex Persian questions that have been decomposed into smaller semantic units in the form of semantic representation. The purpose of this dataset is to facilitate understanding and answering complex Persian questions.

**Data Collection**: The complex questions in this dataset were extracted from multi-entity and multi-hop questions. During the dataset construction process, approximately 15,000 multi-hop complex questions were extracted from the PeCoQ dataset. Out of these, 5,600 questions that were correctly decomposed by the annotators were selected for the final dataset.

**Annotation**: Fourteen Persian language experts, who were well-versed in natural language processing (NLP) and question answering systems, were responsible for annotating the data. Each complex question was decomposed into smaller semantic units. The initial semantic decomposition for each question was performed by two independent annotators, and the quality of these decompositions was then reviewed by another annotator. In cases where discrepancies existed in the annotations, all annotators discussed and finalized the decomposition. The MRDCPQ dataset is divided into three categories: training (80%), testing (10%), and validation (10%).

**Data Generation Process**: To generate the smaller semantic units from complex questions, the information available in the "Named Entities" and "Question Relations" columns was used. Each multi-hop complex question in the PeCoQ dataset includes named entities and relations present in the question. These two aspects were provided as assisting elements to the annotators in separate columns alongside each multi-hop complex question. This allowed the annotators to better decompose the complex questions into simpler semantic units. This information helps annotators perform precise and logical decomposition of the questions. The ultimate goal is to decompose each complex question into smaller units that can be answered sequentially to eventually arrive at the final answer. For instance, in the complex question "How many actors play roles in *Masir Eshgh*?", the semantic decomposition is: "Masir Eshgh; Actors 1#; Number 2#". This structure aids the annotators in breaking down complex questions into simpler and more answerable sub-questions.

In Table 1, an example of a complex question from the PeCoQ dataset, decomposed and presented as part of the MRDCPQ dataset, is provided.

Table 1. A complex Persian question from the PeCoQ dataset decomposed into simplified semantic segments, presented as part of the MRDCPQ dataset.

| Persian Complex Question | Simplified Semantic Decomposed Segments of the Persian Complex Question |
|---|---|
| مساحت کلی کشور فیلم دنیای_آپو چقدر است؟ (What is the total area of the country in the film "The World of Apu"?) | فیلم دنیای_آپو (The World of Apu film) <br> کشور 1# (Country 1#) <br> مساحت کلی 2# (Total area 2#) |

### 3.1.2 The Implementation of Question Decomposition Component

This component was implemented by fine-tuning the mT5 language model [37] on the MRDCPQ dataset. The mT5 model, which supports multiple languages, was chosen due to its ability to perform well on various linguistic tasks across different languages, including Persian. By training the model specifically on the MRDCPQ dataset, the goal was to enable the model to decompose complex Persian questions into simpler, semantic sub-components effectively. The trained model generates outputs that break down multi-step questions into logical segments, allowing for easier processing and reasoning. An example of the output produced by the trained model is shown in Table 2.

Table 2. Examples of Persian Complex Questions and their Semantic Decomposed Segments. This table demonstrates how the mT5 model, trained on the MRDCPQ dataset, decomposes complex Persian questions into simplified, step-by-step semantic segments for easier understanding and processing.

| # | Persian Complex Question | | Semantic Decomposed Segments |
|---|---|---|---|
| 1 | کد منطقه‌ای که شهر تهران در آن قرار دارد، چیست؟ (What is the area code of the city of Tehran?) | #1 | تهران (Tehran) |
| | | #2 | شهر 1# (city) |
| | | #3 | کد منطقه‌ای 2# (Area code) |
| 2 | میانگین دما در محل اقامت مهران_مدیری چقدر است؟ (What is the average temperature at Mehran Modiri's residence?) | #1 | مهران_مدیری (Mehran Modiri) |
| | | #2 | محل اقامت 1# (Residence) |
| | | #3 | میانگین دما 2# (Average temperature) |
| 3 | منطقه ساعت جهانی که کشور کیا_موتورز در آن است چیست؟ (What is the world time zone where Kia Motors is located?) | #1 | کیا_موتورز (Kia Motors) |
| | | #2 | کشور 1# (Country) |
| | | #3 | منطقه ساعت جهانی 2# (World time zone) |
| 4 | شعار کشور محل تولد ایمانوئل_کانت چیست؟ (What is the motto of the country where Immanuel Kant was born?) | #1 | ایمانوئل_کانت (Immanuel Kant) |
| | | #2 | محل تولد 1# (Place of birth) |
| | | #3 | کشور 2# (Country) |
| | | #4 | شعار 3# (Motto) |
| 5 | کشور محل مرگ الکساندر_میخائوفسکی چه سرودی دارد؟ (What is the national anthem of the country where Alexander Mikhailovsky died?) | #1 | الکساندر_میخائوفسکی (Alexander Mikhailovsky) |
| | | #2 | محل مرگ 1# (Place of death) |
| | | #3 | کشور 2# (Country) |
| | | #4 | سرود ملی 3# (National anthem) |

### 3.2 Named Entity Recognition (NER) Component

A complex multi-step question may involve multiple named entities. In the first stage, named entities need to be extracted from the simplified semantic segments derived from the complex question (MRDCPQ). For this purpose, a Named Entity Recognition (NER) module has been developed by fine-tuning the ParsBERT [38] model (Farahani et al., 2020) on the MRDCPQ dataset. The PeCoQ dataset, from which the complex multi-step questions were extracted, includes named entities based on the FarsBase knowledge graph. Therefore, these named entities were applied to the decomposed question segments as the dataset for the NER task to fine-tune the model.

Experimental results indicate that our proposed NER method outperforms existing Persian language tools in extracting named entities from decomposed segments of complex questions. The evaluation of our NER tool is discussed in section 4.

After extracting named entities, the next step is linking them to FarsBase. Following the Zero-shot method introduced by Wu et al [39]. we first extract five candidates from the FarsBase knowledge graph using string similarity. We then map the abstract information of each candidate entity from FarsBase into dense vectors using the ParsBERT [38] model for dense retrieval. Next, the vector of each candidate entity is compared with the vector of the complex question text from which the named entity was extracted. This comparison is conducted using cosine similarity. The

FarsBase candidate entity with the closest vector to the complex question text vector is selected as the final entity. Figure 6 illustrates the process of linking extracted entities from the decomposed question segments.

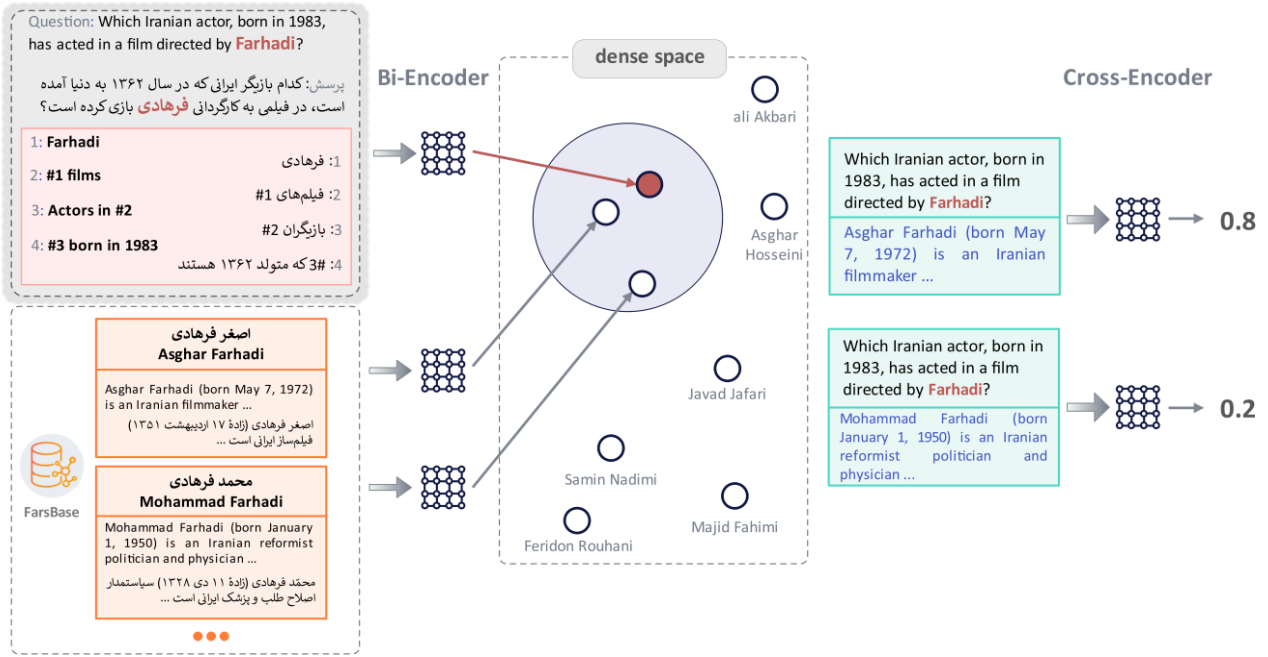

Figure 6. Illustration of our zero-shot entity linking approach for Persian complex questions. The question is first decomposed into semantic components (as shown on the left). These components are encoded into a dense vector space, where named entities are also represented. A nearest neighbor search retrieves potential candidate entities (within the blue circle). The candidates are then passed to a cross-encoder, which compares the original question with entity descriptions to generate a ranking based on their relevance, ultimately selecting the best-matching entity from Farsbase.

*3.3. MRDCPQ to SPARQL Query Generator Component*

In this section, we focus on generating a dataset by leveraging the decomposed components of complex questions (MRDCPQ), which are, in essence, simplified questions normalized to form SPARQL queries. Each MRDCPQ acts as an input, while its corresponding SPARQL query serves as the output. The process of building this dataset, aimed at converting the decomposed parts of complex questions into SPARQL queries, was expedited due to the availability of the Relations key in the PeCoQ dataset.

Dataset Construction Process: For the creation of this dataset, alongside the development of the MRDCPQ dataset, a SPARQL query was generated for each simplified component of the complex question. The creation of these queries relied on the entities extracted from the complex question and the Relations available for each query in the PeCoQ dataset. These extracted entities and relations guided the SPARQL query generation process. Figure 7 illustrates the fine-tuning process of the mT5-based model, which converts decomposed parts of a complex question (MRDCPQ) into executable SPARQL queries.

To enhance the process, the mT5 model was fine-tuned using a custom dataset. This dataset included natural language questions and their corresponding SPARQL queries. By training the model on this specific dataset, the system learned how to map a simplified, structured query into a full SPARQL query. The model was trained to identify key components such as the main entity (e.g., Digikala), relevant relations (e.g., headquarter), and target attributes (e.g., population). This fine-tuning enabled the model to generate SPARQL queries step by step, starting from an entity and progressing through the relations to the final information retrieval. By breaking down each MRDCPQ into its logical steps, the model can automate the query creation process, making the overall conversion from complex natural language queries into SPARQL more efficient.

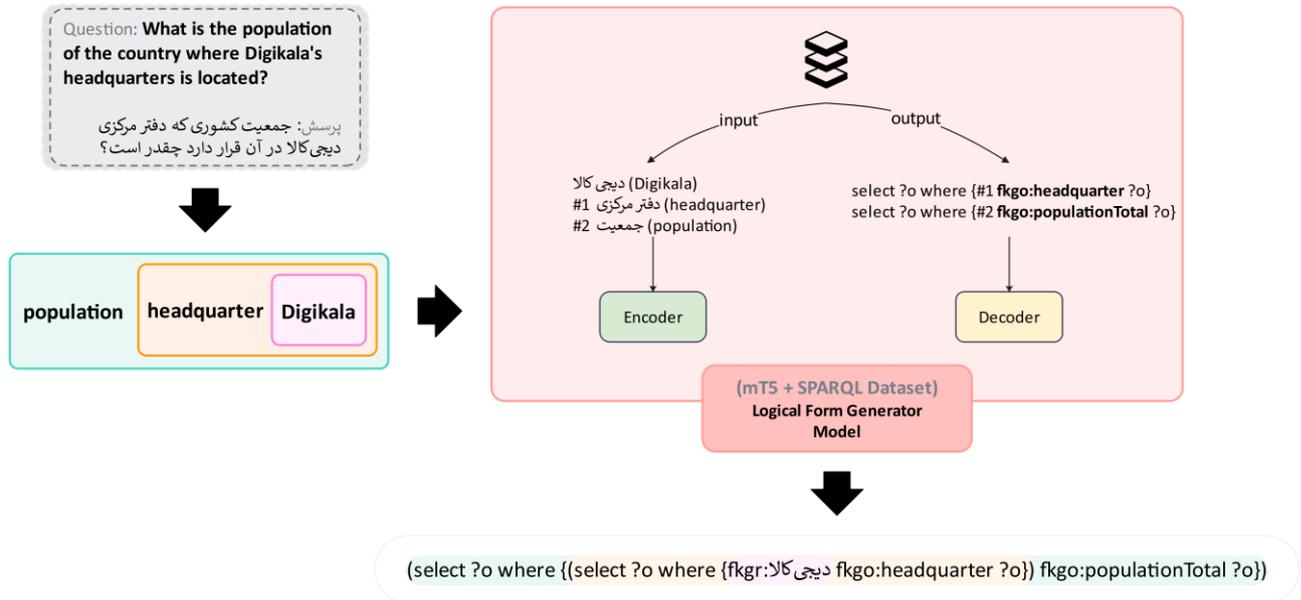

Figure 7. This figure illustrates the fine-tuning process of an mT5-based model for converting decomposed segments of a complex question (MRDCPQ) into executable SPARQL queries. The input is first encoded, where the key components such as "Digikala," "headquarter," and "population" are mapped into SPARQL query forms. The output consists of logical queries that retrieve information, such as the population of the country where Digikala's headquarters is located.

### 3.4. SPARQL Query Execution and Response Composition

This component is responsible for executing the hierarchical SPARQL queries generated in the previous stage on the knowledge graph (KG). Given that the results from each SPARQL query need to be used in the subsequent query, managing this process is another essential task of this component. At the end of the process, the final response is refined and prepared as the answer to the complex question posed.

As illustrated in Figure 8, the breakdown of a complex question into individual simpler components (MRDCPQ) helps structure the query-building process. Each parsed component contributes to a distinct SPARQL query step. For example, the figure 8 demonstrates how the question "What is the population of the country where Digikala's headquarters is located?" is split into three parts: "Digikala," "headquarters," and "population," resulting in successive SPARQL queries to gather the required information from the knowledge graph.

The component consists of two subcomponents:

1. **Execution of SPARQL Queries on the Knowledge Graph**: Handles the execution of each query step-by-step, ensuring that results from one query feed into the next.
2. **Final Response Preparation**: Formats and finalizes the output, ensuring it is well-structured for the original complex question.
3. 

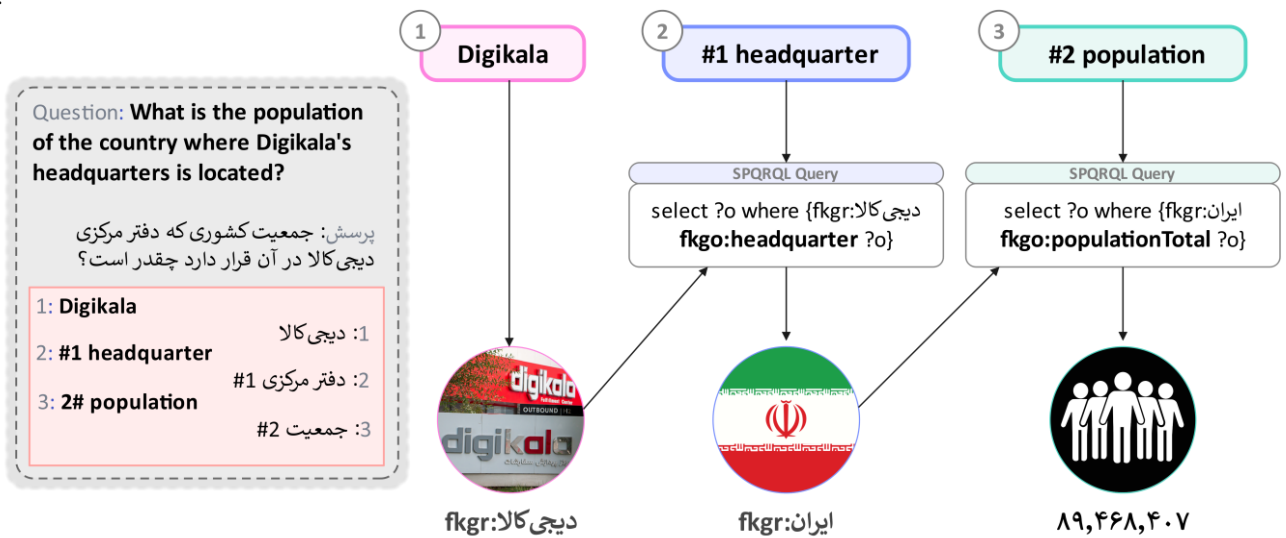

Figure 8. The process of converting a complex question into simpler components and building successive SPARQL queries for each component, as shown for the example question about the population of the country where Digikala's headquarters is located.

## 4. Evaluation

To evaluate the proposed method, it is necessary to assess the constituent components and the efficiency of each, which will be detailed subsequently.

*4.1 Question Decomposition Component*

In this research, the evaluation of the model for decomposing complex questions into simpler questions based on the **Task Decomposition Accuracy (TDA)** metric yielded a score of **77.61%**. TDA is particularly well-suited for evaluating the precision of breaking down complex questions into simpler ones. This metric is highly effective in assessing the core task at hand, which is decomposing complex questions into simpler, manageable steps. The focus of TDA is on evaluating the accuracy of splitting complex tasks into simpler subtasks.

The **Task Decomposition Accuracy (TDA)** metric measures whether the decomposition of complex questions into simpler questions has been carried out correctly. In other words, it examines whether the extracted steps are logically correct, precise, and arranged in the proper sequence.

Key criteria assessed using the TDA method include:
- **Semantic Segment Accuracy**: Each part of the question must be accurately transformed into a simpler step.
- **Correct Sequencing**: The steps should be logically connected and follow one another in a step-by-step manner.
- **Complete Coverage of the Question**: The decomposed questions must cover the entire meaning of the complex question.

*4.2 Named Entity Recognition (NER) Component*

Table 3 demonstrates that fine-tuning ParsBERT on the MRDCPQ dataset significantly improves its performance in the Named Entity Recognition (NER) task compared to ParsBERT models trained on the Arman and PeCoQ datasets. The fine-tuning process allows ParsBERT to better capture domain-specific nuances, particularly when handling multi-hop complex questions, thereby enhancing its ability to recognize named entities more accurately. The comparison highlights the importance of task-specific training data in improving the accuracy of NER models, especially in complex question-answering systems based on knowledge graphs.

Table 3. Accuracy of Named Entity Recognition (NER) models on the MRDCPQ test set. The results show that fine-tuning ParsBERT on the MRDCPQ dataset yields the highest accuracy, significantly outperforming both the model trained on the ARMAN dataset and the fine-tuned ParsBERT on PeCoQ.

| Named Entity Recognition (NER) Component | Accuracy |
|---|---|
| ParsBERT-ARMAN | 51.74% |
| Fine-Tuned ParsBERT-PeCoQ | 96.11% |
| **Fine-Tuned ParsBERT-MRDCPQ** | **99.16%** |

*4.3. MRDCPQ to SPARQL Query Generator Component*

In the process of generating SPARQL queries from the simplified questions (MRDCPQ), which were decomposed by the "Question Decomposition Component" in the previous stage, we utilized a fine-tuned mT5 model trained on a prepared dataset. The evaluation of this step shows that the model effectively transformed the simplified questions into their corresponding SPARQL queries, achieving an F1 Score of 82.35%. This result highlights the model's robust performance in converting natural language questions into executable logical forms on the knowledge graph.

*4.4. The Component of Extracting Answers from the Knowledge Graph*

In order to assess the efficiency of the proposed method, the results obtained from the Component of Extracting Answers from the Knowledge Graph in our approach were compared to the only existing similar work, namely the research by Etezadi et al. [23], which includes the PeCoQ dataset and a proposed evaluation method for it. As illustrated in Table 4 and Figure 9, the findings demonstrate that our proposed method achieves significant improvements across key metrics: precision by 13.12%, recall by 11.96%, F1 score by 12.57%, and accuracy by 12.06% when compared to Etezadi's method.

Table 4. The obtained results from the evaluation of the proposed method compared to the similar work on the test dataset of PeCoQ.

|  | Precision | Recall | F1 Score | Accuracy |
|---|---|---|---|---|
| **Our Method** | 84.36% | 68.41% | 75.55% | **74.81%** |
| Etezadi's Method | 71.24% | 56.45% | 62.98% | 62.75% |

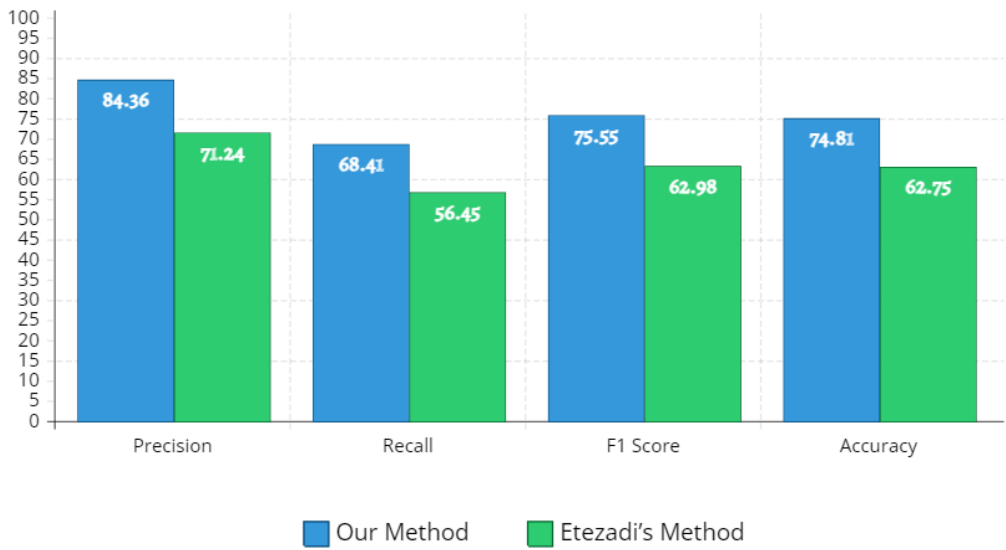

Figure 9. The obtained results in the proposed method compared to the similar work across different metrics.

## 5. Conclusion and Future Works

In this paper, we proposed a novel approach to address the challenge of answering complex natural language questions by leveraging a Persian knowledge graph. Our method consists of four main components: question parsing, named entity recognition, conversion of complex questions into SPARQL queries, and answer extraction from the knowledge graph. These components work together to accurately respond to user queries. A significant challenge we focused on was the transformation of complex questions into SPARQL queries, which was achieved by breaking down complex questions into simpler sub-questions, converting each into a SPARQL query, and then merging the results. The proposed approach was tested against existing methods using the PeCoQ dataset, and the results demonstrated its superiority in various evaluation metrics.

In future work, we aim to explore the following directions to further enhance the system:
1. **Expanding the dataset for question decomposition:** Increasing the dataset size and diversity to include a broader spectrum of complex questions across different domains.
2. **Improving language models:** Incorporating advanced language models, such as large-scale models, to increase system accuracy and efficiency.
3. **Exploring multimodal learning:** Integrating textual and visual data to enable the system to handle multimedia-based queries.

These directions could significantly improve the performance of question-answering systems and lead to a better user experience when dealing with complex queries.

## Authors' Profiles

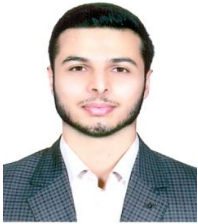

**Arash Ghafouri:** He is a PhD candidate in Computer Engineering at the Iran University of Science and Technology. He is active in the field of natural language processing, information retrieval, and their applications in other areas such as search engines and chatbots. He received his Master's degree in Computer Engineering from the Iran University of Science and Technology in 2014. As an active researcher in the field of natural language processing and information retrieval, he used novel natural language processing methods and advanced machine learning techniques to solve problems related to these fields.
Email: aghafuri@comp.iust.ac.ir

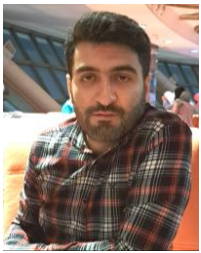

**Mahdi Firouzmandi:** He received the bachelor's degree in computer engineering from the Department of Computer Engineering, Iran University of Science and Technology, Tehran, Iran, in 2013, where he is currently pursuing the master's degree. His research activity concerns artificial intelligence and machine learning, focusing on question answering system and related natural language processing (NLP) tasks. Other research interests involve deep learning and its application in other research field
Email: firouzmandiii@gmail.com

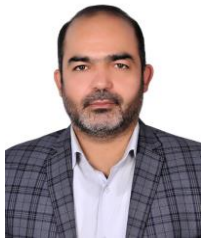

**Hassan Naderi:** He is an associate professor at the Iran University of Science and Technology. He received the M.S. degree in computer engineering from the Sharif University of Technology, Tehran, in 2001, and the Ph.D. degree in information technology from the Institut National des Sciences Appliquées de Lyon, France, in 2006. He leads the Data Science and Technology Laboratory (DSTL), which researches various areas in data science and its application application in other research field.
Email: naderi@iust.ac.ir